# IMPACT VITALITY - A MEASURE FOR EXCELLENT SCIENTISTS

Nadine Rons*

Vrije Universiteit Brussel (VUB), R&D dept., Pleinlaan 2, B-1050 Brussels, Belgium.
E-mail: nrons@vub.ac.be;
Phone: +32 (0)2 629.21.79; Fax: +32 (0)2 629.36.40

Lucy Amez

Vrije Universiteit Brussel (VUB) and Policy Research Centre for R&D Indicators VUB (SOOI-VUB), Pleinlaan 2, B-1050 Brussels, Belgium.
E-mail: Lucy.Amez@vub.ac.be

* Corresponding author

## 1 Background

In many countries and at European level, research policy increasingly focuses on 'excellent' researchers. The concept of excellence however is complex and multidimensional. For individual scholars it involves talents for innovative knowledge creation and successful transmission to peers, as well as management capacities. Excellence is also a comparative concept, implying the ability to surpass others [TIJSSEN, 2003]. Grants are in general awarded based on assessments by expert committees. While peer review is a widely accepted practice, it nevertheless is also subject to criticism. At higher aggregation levels, peer assessments are often supported by quantitative measures. At individual level, most of these measures are much less appropriate and there is a need for new, dedicated indicators.

## 2 Research question

The central question is how to define excellent scientists, so that they can be recognized among very good researchers in a way suited to a policy tool. We apply this question to publication activity. A suitable indicator further needs to cope with several prerequisites and challenges. Briefly, it needs to conceptually represent capacities sought, be validated preferably by correlations with peers' appreciations, minimize influence of occasional outlyers and faulty data, avoid bias, reflect recent performance and be easy to calculate to serve policy needs, and be hard to manipulate not to perturb a good scientific publication culture.

## 3 Methodology

From excellence programmes, qualitative appreciations were selected which distinguish excellence from lower performance levels and which apply to publication and citation behaviour. After translation into measurable quantities, an indicator was built considering the prerequisites and challenges involved.

## 4 Results

The derived description of excellence points towards the extent to which one's work is cited, in particular recently and increasingly:

*An excellent researcher is prominently present in the field, continuously publishing new knowledge and ideas over a longer period of time. As an established reference in the field, his/her contributions are eagerly followed by colleagues and his/her ideas are picked up fast in their further research. As such, he or she is a central figure in a strong research dynamic, at the level of the researcher's own research team as well as for the research area as a whole, increasing both volume and impact of research in the field.*

Inspired by the new vitality measure announced by KLAVANS & BOYACK [2006] for national and institutional comparison, we propose an *impact* vitality measure (IV), not modulated by age of references, but by age of publications citing a scientist's work. For the impact vitality $IV(y_1, n)$ in year $y_1$, with a window starting n years back in $y_n$, citing publications $P(y_i)$ published in year $y_i$ receive a lower weight when having a higher age i. Normalization yields a value larger than 1 when citing publications increase with time, and smaller than 1 when they decrease.

$$IV(y_1, n) = [n \, (\sum_{i=1 \to n} P(y_i) / i) / \sum_{i=1 \to n} P(y_i) - 1] / [\sum_{i=1 \to n} 1/i - 1]$$

with $n > 1$, $y_{i+1} = y_i - 1$ and $\sum_{i=1 \to n} P(y_i) > 0$

Options for calculations in subsequent years include a moving time window of fixed length, or a growing window starting from a fixed year, for example the year $y_{PhD}$ the scientist obtained a PhD:

$$IV_{PhD}(y_1) = IV(y_1, n_{PhD}), \text{ where } n_{PhD} = y_1 - y_{PhD} + 1, \text{ the age in years of the PhD.}$$

A first test sample of applicants to open calls for senior research fellowships (excluding candidates rejected for formal reasons), showed that selected applicants all had $IV_{PhD}$-values $\geq 1$ for all years since their PhD, and that none were selected who had an $IV_{PhD}$-value $<1$ for one or more years.

**Table 1** Peer review based selection vs. Impact Vitality: first test results

|  | $IV_{PhD} \geq 1$ for all years | $IV_{PhD} < 1$ for one or more years |
|---|---|---|
| Selected | 5 | — |
| Not selected | 4 | 4 |

Source: VUB Research Fellowship calls 2000-2006, applications in predefined research themes, excluding themes in Social Sciences and Humanities. Impact Vitality calculated using the Web of Science.

## 5 Conclusion and further research

A novel indicator is proposed to help identify excellent scientists, reflecting a sustained increase of publications that cite their work. It is relatively easy to calculate and hard to manipulate and has a limited sensitivity to outlyers in citation counts and to faults in references. Further features are a scope broader than the scientist's indexed publications and independence regarding size and citation culture of the research community. While a first test of limited size looks promising, further research is necessary on larger samples and for different indicator variants before any use can be recommended within a set of indicators for the assessment of individual scientists.